\documentstyle[12pt,epsf,epsfig,wrapfig2]{article}

\newcommand{\be}{\begin{eqnarray}}
\newcommand{\ee}{\end{eqnarray}}

\begin{document}

\begin{center}

{\bfseries Charmonium polarization in $e^+e^-$ and heavy ion
collisions}

\vskip 5mm

B.L. Ioffe

\vskip 5mm

{\small

{\it  Institute of Theoretical and Experimental Physics, B.
Cheremushkinskaya 25,\\
 Moscow 117218, Russia, E-mail: ioffe@vitep1.itep.ru}
 }

\end{center}

\begin{abstract}

In $e^+e^-$ annihilation at $\sqrt{s}=10.6$ GeV Belle
Collaboration found, that $J/\psi$ mesons are predominantly
produced in association with an extra $\overline{c}c$ pair. The
possible mechanisms of $J/\psi$ production are discussed and the
probability of the associate production of $\overline{c}c$ pair is
estimated. The choice between these mechanisms can be done by
measuring $J/\psi$ polarization. It is suggested, that in case of
heavy ion collisions one may expect remarkable  transverse
polarization of produced $J/\psi$, if quark-gluon plasma if
formed. The measurement of asymmetry of $e^+e^-(\mu^+\mu^-)$
angular distribution in $J/\psi\to e^+e^-(\mu^+\mu^-)$ decay is a
useful tool for detection of quark-gluon plasma formation in heavy
ion collisions.

\end{abstract}

{\bf 1.} Recently, Belle Collaboration reported the measurement of
$J/\psi$ production in $e^+e^-$ annihilation at $\sqrt{s}=10.6$
GeV \cite{Abe:2002rb}. It was found that the cross section of
$J/\psi$ production significantly exceeds theoretical expectations
based on the Color Singlet Model (CSM) \cite{CL} and
non-relativistic QCD (NRQCD) \cite{SB,YQC}. Even more
surprisingly, it was found that most of the observed $J/\psi$'s
were accompanied by an extra $\bar{c}c$ pair, with $\sigma(e^+e^-
\to J/\psi \bar{c}c) / \sigma(e^+e^- \to J/\psi X) =
0.59^{+0.15}_{-0.13}\pm 0.12$. This ratio exceeds the existing
theoretical predictions by about an order of magnitude. The
problem has recently attracted attention of many theorists -- see,
e.g. \cite{Brodsky:2003hv}-\cite{AK}.

 Let us consider the
diagrams of $J/\psi$ production in $e^+e^-$ annihilation. There
are three types of such diagrams. In the diagrams of the first
type (Fig.1a) $J/\psi$  is formed by the fusion of $\bar{c}c$ pair
produced by the initial virtual photon. This means that $\bar{c}c$
quarks, which initially were moving in the opposite directions,
turn around and have almost equal and parallel momenta in the
final state. The momentum conservation is ensured by the radiated
gluons which may produce light ($q=u,d,s$) $\bar{q}q$ or charmed
$\bar{c}c$ pairs. The diagrams of the second type (Fig.1b)
correspond to the fragmentation of $c$ (or $\bar{c}$) into
$J/\psi$. This process requires the production of an additional
$\bar{c}c$ pair by emission of at  least one gluon by initial $c$
or $\bar{c}$. Finally, the third type of processes are described
by the diagrams in which the initial virtual photon creates a pair
of light quarks and the $J/\psi$ is formed from the $\bar{c}c$
pair produced by an exchange of gluons (Fig.1c). Evidently, the
diagram of Fig.1c is suppressed in comparison to diagrams of
Figs.1a,b by a factor of $\alpha_s(m_c)$ and will be disregarded
in what follows.

On general grounds one may expect that the diagrams of Fig.1b
dominate at very high energies of the colliding $e^+e^-$, when the
ladder of $\bar{c}c$ pairs is formed and the process may be
described by Regge theory (such process was considered by Kaidalov
\cite{AK}). It is easy to estimate the energies starting from
which one may expect the dominance of the diagrams of Fig.1b. Let
$p_c$ be the momentum of the $c-$quark (in the $e^+e^-$ c.m.
system) fragmenting into $J/\psi$. Then the minimal value of the
recoil momentum $q$, corresponding to the forward production, is
equal to
\be
q \simeq {M_{J/\psi}^2 - m_c^2 \over 2 p_c}.
\ee
By requiring $q$ to be at least as small as $q \simeq 0.5$ GeV
(typical for Regge asymptotics), we get
$p_c > 10 $ GeV, i.e. $\sqrt{s} > 20$ GeV. In fact one may expect
that the energy should be $\sqrt{s} > 50$ GeV.
\begin{wrapfigure}{R}{8cm}
\mbox{\epsfig{figure=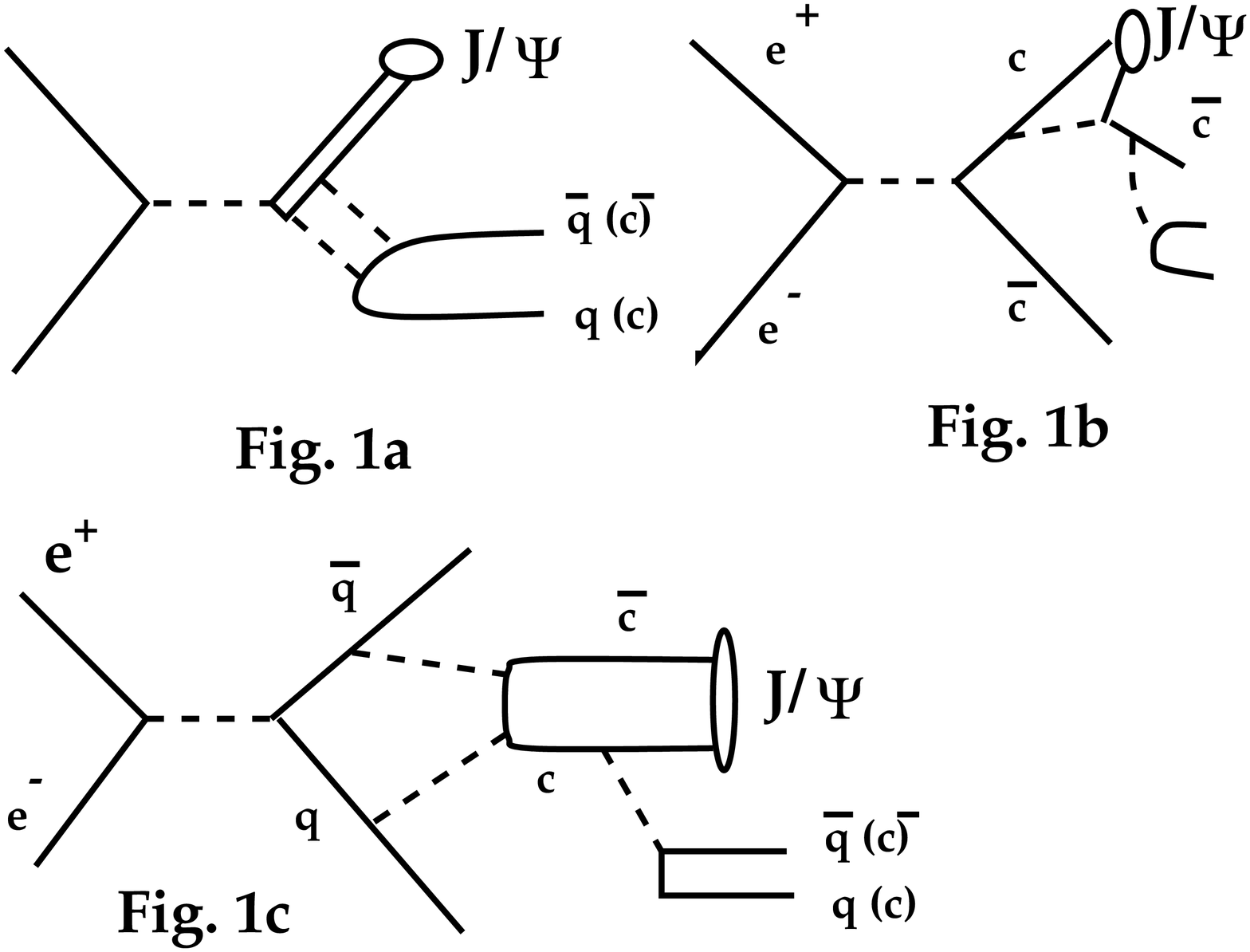,width=7.8cm,height=6cm}} {\small{\bf
Figure 1.} The production of $J/\psi$ in $e^+e^-$ annihilation.}
\end{wrapfigure}
 Probably, the mechanism of Fig.1b cannot be the dominating one
at $\sqrt{s}=10.6$ GeV. The more suitable candidate for
description of $J/\psi$ production in $e^+e^-$-annihilation at
this energy is Fig.1a mechanism.

Belle Collaboration \cite{Abe:2002rb} measured the distribution of
events as a function of the mass of the system recoiling against
the $J/\psi$. The recoil mass was defined as
\be
M_{recoil}=\sqrt{(\sqrt{s}-E_{J/\psi}^*)^2 - p_{J/\psi}^{*2}} \ee
where $E_{J/\psi}^*$ and $p_{J/\psi}^*$ are the energy and
momentum of $J/\psi$ in the c.m. frame. It was found that
$M_{recoil}$ concentrates in the domain $M_{recoil} \geq 5$ GeV,
with almost no events below $2.8$ GeV \cite{SB}. This invariant
mass is sufficiently larger than the threshold for the production
of an additional $\bar{c}c$ pair. At such large $M_{recoil}$ one
can expect, that the perturbative theory is valid in the
production of additional to $J/\psi$ hadrons and the mass of $c$
quarks may be neglected. If all additional quarks are produced
incoherently, then one would expect
\be
R\equiv \frac{\sigma(J/\psi \overline{c}c)}{\sigma(J/\psi X)}
\approx \frac{1}{4}\ee for
$X=\overline{c}c+\overline{q}q,~(q=u,d,s)$. If they are produced
coherently, then light quarks are in the $SU(3)$ singlet state $
|X_{light} \rangle =  |\bar{u}u + \bar{d}d + \bar{s}s
\rangle\sqrt{3}$.  Therefore, light and charm quarks should be
produced in association with the $J/\psi$ with equal
probabilities, and we find
\be
 \sigma(e^+e^- \to J/\psi \ \bar{c}c) /
\sigma(e^+e^- \to J/\psi \ X) = \frac{1}{2}\ee Probably, the true
answer is somewhere between the estimations (3) and (4).

In fragmentation mechanism (Fig.1b) we have evidently $R=1$.

Let us now turn to the discussion of $J/\psi$ polarization. Since
the photon in $e^+e^-$ annihilation is mostly transverse, it has
helicity $\lambda = \pm 1$. Therefore the $\bar{c}c$ produced by
the photon should have opposite helicities of $\lambda_c= + 1/2,
\lambda_{\bar{c}} = -1/2$ or $\lambda_c= - 1/2, \lambda_{\bar{c}}
= + 1/2$. Initially, heavy quark and antiquark move in the
opposite directions in the center of mass system of $e^+e^-$
annihilation with the velocities $v = \sqrt{1-4m_c^2/s}$ which are
close to $v \simeq 1$ at $\sqrt{s} = 10.6$ GeV. However, to become
bound in the $J/\psi$ (or any other bound state of charmonium), at
least one of the quarks has to change the direction of its
momentum by radiating gluons (and extra quark--antiquark pair(s))
since the relative velocity of heavy quarks in a bound state
should be small. Since in QCD  the helicity of the quark is
conserved, a change in the direction of its momentum should be
accompanied by the spin flip. We thus come to the conclusion that
in the case of $J/\psi$ produced at high momentum, the total spin
of $J/\psi$ should have zero projection on its direction of
motion, which corresponds to the longitudinal polarization.

This is in agreement with the experimental result of the BaBar
Collaboration \cite{Aubert:2001pd}, which states that the angular
distribution of positively charged lepton decay product with
respect to the direction of $J/\psi$ measured in the CM frame is
$W(\theta) \sim 1 + \alpha \ cos^2 \theta$ with $\alpha = -0.46
\pm 0.21$ for CM momentum $p^* < 3.5$ GeV, and $\alpha = -0.80 \pm
0.09$ for CM momentum $p^* > 3.5$ GeV. (In this distribution,
$\alpha = -1$ corresponds to longitudinal polarization,
 $\alpha = +1$ to transverse, and $\alpha = 0$ indicates no
polarization).

As is clear from Fig.1b one of the $c$ quarks has helicity, say,
$\lambda = +1/2$; the helicity of the other quark created from the
vacuum is uniformly distributed. Therefore the mean value of
$\alpha$ is equal to zero (it is easy to see that the cases of
$\alpha=+1$ and $\alpha=-1$ have equal probabilities), and the
produced $J/\psi$ is unpolarized. The conclusion is that the
precise measurement of $J/\psi$ polarization can distinguish among
two mentioned above production mechanisms.

\bigskip
{\bf 2.} The possibility to form quark--gluon plasma in heavy ion
collisions is an intriguing problem of strong interaction physics.
To establish the formation of plasma, a number of signatures were
proposed; here we will concentrate on heavy quarkonia. Suppression
of heavy quarkonium states has been suggested long time ago by
Matsui and Satz \cite{matsui} as a signature of the deconfinement
phase transition in heavy ion collisions. Their, by now
well--known, idea is that the Debye screening of the gluon
exchanges will make the binding of heavy quarks into the bound
states impossible or unlikely once a sufficiently high temperature
is reached. The lack of quarkonium states would thus signal
deconfinement; this effect was indeed observed and studied in
detail at CERN SPS by the NA38 \cite{NA38} and NA50 Collaborations
\cite{NA50}. The results on $J/\psi$ production at RHIC have
recently been presented by the PHENIX Collaboration \cite{phenix}.
The observations of quarkonium suppression have been interpreted
as a signal of quark--gluon plasma formation \cite{qgp}. However,
different conclusions were reached in \cite{comov}, where it was
argued that the effect may arise due to quarkonium collisions with
the comoving hadrons. Additional tests of the quark--gluon plasma
formation could help to clarify the situation.

I would like to present here the idea: to use the polarization of
$J/\psi$, produced in heavy ion collisions for diagnostics of
quark-gluon plasma.

Let me first formulate what I mean by the quark--gluon plasma,
since different definitions sometimes may result in
misunderstanding. I define the quark--gluon plasma as a gas of
quarks and gluons in which the interactions can be described by
perturbative QCD and non--perturbative effects are either absent
or can be neglected. It is no need to specify the properties of
this state of matter in more detail to develop the idea.

It is well--known that the description of the data on heavy
quarkonium production within the framework of perturbative QCD
(pQCD) meets with siginificant difficulties. Both the absolute
values of the measured production cross sections of hidden heavy
flavor states and the relative abundances of different quarkonia
are not described well within the perturbative framework, but
perhaps the most spectacular failure of pQCD is the polarization
of the produced quarkonia. Even an extension of a perturbative
approach based on non--relativistic QCD \cite{BBL}, which allows
certain non--perturbative physics, does not allow to explain the
polarization measurements \cite{CDF}.

Let us illustrate this idea in more detail using the example of
$J/\psi$ polarization. There are two mechanisms of $J/\psi$
production in hadron collisions -- direct, when $J/\psi$ is
produced by perturbative and non--perturbative interactions of
gluons and quarks, and cascade, when $J/\psi$ is created as a
result of decays of C--even $\bar{c}c$ states, $\chi_c \to J/\psi
+ \gamma$. In quark--gluon plasma, the cascade production
mechanism should be at least as important as direct production.
Indeed, in the lowest order of perturbation theory,  $J/\psi$ is
produced by the three gluon fusion or by two gluon fusion followed
by the gluon emission off the $\bar{c}c$ system. In both cases the
probability of   $J/\psi$ production is proportional to
$\alpha_s^3(m_c)$. The probability of $\chi_c^{0,2}$ production is
proportional to $\alpha_s^2(m_c)$, i.e. it is of lower order in
$\alpha_s$, which however is largely compensated by the branching
ratio $B(\chi_2 \to J/\psi + \gamma) \simeq 20 \%$ for the
$J/\psi$ production.

In ref \cite{VHBT}  $J/\psi$ production cross section in $\pi N$
interactions was calculated in perturbation theory. The
contributions from various sources to the $J/\psi$ production in
$\pi^- N$ collisions at the incident energy of $185$ and $300$ GeV
are shown in Table 1 (the data are from \cite{lemo}).


\begin{center}
 {\bf Table 1}

\vspace{3mm} {\footnotesize
\begin{tabular}{|c|c|c|c|c|c|c|c|} \hline &&& &
\multicolumn{4}{c|}{$\sigma(J/\psi),nb$}
\\ \cline{5-8}
& $\sigma(\chi_2),nb$ & {\normalsize
$\frac{\sigma_{dir}(J/\psi)}{\sigma(\chi_2\to J/\psi)}$} &
{\normalsize $ \frac{\sigma(\chi_1)}{\sigma(\chi_2)}$} & dir  &
 $\chi_2\to
J/\psi\gamma$ &  $\chi_1\to J/\psi\gamma$ & Total
\\ \hline Exper. & $188\pm 30\pm21$ &
$ 0.54\pm 0.1\pm 0.1$ & $0.70\pm 0.15$ & 102 & 35 & 42 & 180
\\
 Theory & 78 & ~0.17 & 0.067& 13.2 & 14.7 &1.6 & 29.5 \\ \hline
\end{tabular}}
\end{center}

As is clear from Table 1 the perturbative calculations of $J/\psi$
production in $\pi N$ collisions disagree with the data by a
factor 6: the  nonperturbative effects are dominant.

Let us now turn to $J/\psi$ polarization as reconstructed from the
angular distributions of electrons (muons) from the $J/\psi \to
e^+ e^- (\mu^+ \mu^-)$ decays. Generally the electron (muon)
distribution has the form
\be
W(\theta) \sim 1 + \alpha \ cos^2 \theta, \label{dist} \ee where
$\theta$ is the emission angle of $e^+$ (or $\mu^+$) relative to
the direction of $J/\psi$ motion in its rest frame; at small
$p_t$, this direction coincides with the direction of the beam.
The value $\alpha = 1$ corresponds to the transverse polarization,
$\alpha = -1$ -- to the longitudinal polarization, and $\alpha =
0$ to unpolarized $J/\psi$. In perturbation theory, in the case
when  $J/\psi$ is produced through the
  $\chi_2 \to J/\psi + \gamma$ decay, the coefficient $\alpha$ in Eq.
(\ref{dist}) is determined unambiguously (at small $p_t$): $\alpha
= 1$ \cite{BI}. This comes from the fact that $\chi_2$ is produced
by two--gluon fusion, $gg \to \chi_2$, for which the effective
interaction is $f_{\mu\nu} \Theta_{\mu\nu}$, where
$\Theta_{\mu\nu}$ is the energy--mometum tensor of the gluon field
and $f_{\mu\nu}$ is the wave function of $\chi_2$. Since
$\Theta_{\mu\nu}$ has only $J_z = \pm 2$ spin projections on the
direction of gluon momenta (indeed,  $\Theta_{\mu\nu}$ may be
considered as a source of the graviton field), the same spin
projections has the $\chi_2$. As a result, $J/\psi$ produced via
$\chi_2$ decay is transversely polarized, $J_z = \pm 1$ and thus
$\alpha =1$.

This conclusion is somewhat modified when the initial transverse
momenta of the gluons are taken into account. This reduces the
value of $\alpha$ to \cite{BI}
\be
\alpha \longrightarrow \alpha' = \alpha {(1 - {3 \over 2}\
\theta_0^2)\over 1 + \alpha\ \theta_0^2/2}, \label{pt} \ee where
$\theta_0^2 \sim 4 \langle p_t^2 \rangle /M_{\chi}^2$.
 For $p_t \sim 1$ GeV, the formula Eq.(\ref{pt}) yields
a reduction of polarization down to $\alpha \simeq 0.5$; still,
this value corresponds to a significant transverse polarization.

The asymmetry coefficient $\alpha$ was also computed for the
directly produced $J/\psi$ and for the production via the $\chi_1$
decay \cite{VHBT}. The results are $\alpha_{dir} \simeq 0.25$ for
direct production and $\alpha_{\chi_1} \simeq -0.15$ for the
production via $\chi_1$ decay (except the forward region of $x_F >
0.8$, where both $\alpha_{dir}$ and $\alpha_{\chi_1}$ begin to
increase). After summing all channels of $J/\psi$ production it
was found \cite{VHBT} that $\alpha_{tot}^{pert} \simeq 0.5$.
Experimentally \cite{data}, no sizable polarization in the entire
range of $x_F$ was observed, $\alpha \simeq 0$ (there is however
an indication that at very large $x_F$ $\alpha$ becomes negative).
This disagreement between theory and experiment demonstrates again
that the production mechanism of $J/\psi$, and possibly $\chi_1$
and $\chi_2$ in hadronic collisions is essentially
non--perturbative.

Let is now dwell upon the $J/\psi$ production in heavy ion
collisions. Let us assume that at sufficiently high collision
energy the quark--gluon plasma is formed. Due to the arguments
presented above, the formation of quarkonia will thus take place
in the plasma (this will of course result in the suppression of
the formation probability \cite{matsui}). The non--perturbative
effects should thus be absent (or small), and we are left only
with the perturbative mechanism. Then, according to the third row
of Table 1, about one half of $J/\psi$'s will be produced directly
and another one half via $\chi_2 \to J/\psi + \gamma$. (The
approximate equality of these contributions stems from the fact
that the extra power of $\alpha_s$ in the direct production cross
section is compensated by a relatively small branching ratio --
about $20 \%$ -- of the $\chi_2 \to J/\psi + \gamma$ decay.) We
thus expect that the asymmetry coefficient of the electron (muon)
angular distribution in the $J/\psi \to e^+e^-(\mu^+\mu^-)$ decay
in the case of quark--gluon plasma formation will increase from
zero to about (at $p_t = 0$) $\alpha \simeq 0.6$. The account of
the initial transverse momentum distribution of gluons as
discussed above reduces asymmetry coefficient to
\be
\alpha \simeq 0.35 \div 0.4. \label{finres} \ee

\vskip0.3cm

{\bf Conclusion.} In case of quark-gluon plasma formation in heavy
ion collisions, one may expect an essential increase of $J\psi$
polarization in comparison with that in hadronic collisions.
Therefore, the measurement of electron (muon) angular asymmetry of
$J/\psi \to e^+e^- (\mu^+\mu^-)$ decay is an effective tool of
detection of quark-gluon plasma formation in heavy ion collisions.

The content of my talk is based of the papers by D.Kharzeev and
myself \cite{BLKh},\cite{BLDE}. This work was partially supported by INTAS grant 2000-587
 and RFBR grant 03-02-16209.



\begin{thebibliography}{99}
\bibitem{Abe:2002rb}
K.~Abe {\it et al.}  [Belle Collaboration],
Phys.\ Rev.\ Lett.\  {\bf 89}, 142001 (2002)

\bibitem{CL}
P.~L.~Cho and A.~K.~Leibovich,
Phys.\ Rev.\ D {\bf 54}, 6690 (1996)

\bibitem{SB}
S.~Baek, P.~Ko, J.~Lee and H.~S.~Song,
J.\ Korean Phys.\ Soc.\  {\bf 33}, 97 (1998)


\bibitem{YQC}
F.~Yuan, C.~F.~Qiao and K.~T.~Chao,
Phys.\ Rev.\ D {\bf 56}, 321 (1997)

\bibitem{Brodsky:2003hv}
S.~J.~Brodsky, A.~S.~Goldhaber and J.~Lee,
hep-ph/0305269.

\bibitem{Luchinsky:2003ej}
A.~V.~Luchinsky,
hep-ph/0305253.

\bibitem{Hagiwara:2003cw}
K.~Hagiwara, E.~Kou and C.~F.~Qiao,
hep-ph/0305102.

\bibitem{Berezhnoy:2003hz}
A.~V.~Berezhnoy and A.~K.~Likhoded,
hep-ph/0303145.

\bibitem{Liu:2003jj}
K.~Y.~Liu, Z.~G.~He and K.~T.~Chao,
hep-ph/0301218.

\bibitem{Bodwin:2002kk}
G.~T.~Bodwin, J.~Lee and E.~Braaten,
Phys.\ Rev.\ D {\bf 67}, 054023 (2003).

\bibitem{Bodwin:2002fk}
G.~T.~Bodwin, J.~Lee and E.~Braaten,
Phys.\ Rev.\ Lett.\  {\bf 90}, 162001 (2003).

\bibitem{AK}
A.~B.~Kaidalov,
hep-ph/0301246.


\bibitem{Aubert:2001pd}
B.~Aubert {\it et al.}  [BABAR Collaboration],
Phys.\ Rev.\ Lett.\  {\bf 87}, 162002 (2001).


\bibitem{matsui}
T.~Matsui and H.Satz, Phys.Lett. B{\bf 178}, 416 (1986).

\bibitem{NA38}
C.~Baglin {\it et al.},
Phys.\ Lett.\ B {\bf 220}, 471 (1989);  B {\bf 255}, 459 (1991);

M.~C.~Abreu {\it et al.},
Phys.\ Lett.\ B {\bf 449}, 128 (1999).

\bibitem{NA50}
M.~C.~Abreu {\it et al.},
Phys.\ Lett.\ B {\bf 410}, 337 (1997);
 B {\bf 477}, 28 (2000);
 B {\bf 521}, 195 (2001).

\bibitem{phenix}
A.~D.~Frawley et al. [PHENIX Collaboration],
nucl-ex/0210013.


\bibitem{qgp}
J.~P.~Blaizot and J.~Y.~Ollitrault,
Phys.\ Rev.\ Lett.\  {\bf 77}, 1703 (1996);

C.~Y.~Wong,
Nucl.\ Phys.\ A {\bf 610}, 434C (1996);

D.~Kharzeev, C.~Lourenco, M.~Nardi and H.~Satz,
Z.\ Phys.\ C {\bf 74}, 307 (1997);

D.~Kharzeev,
Nucl.\ Phys.\ A {\bf 610}, 418C (1996).

\bibitem{comov}


A.~Capella, A.~Kaidalov, A.~Kouider Akil and C.~Gerschel,
Phys.\ Lett.\ B {\bf 393}, 431 (1997);

S.~Gavin and R.~Vogt,
Nucl.\ Phys.\ A {\bf 610}, 442C (1996);

A.~Capella, E.~G.~Ferreiro and A.~B.~Kaidalov,
Phys.\ Rev.\ Lett.\  {\bf 85}, 2080 (2000);

N.~Armesto, A.~Capella, E.~G.~Ferreiro, A.~Kaidalov and D.~Sousa,
Nucl.\ Phys.\ A {\bf 698}, 583 (2002).


\bibitem{BBL}
G.~T.~Bodwin, E.~Braaten and G.~P.~Lepage,
Phys.\ Rev.\ D {\bf 51}, 1125 (1995).

\bibitem{CDF}
T.~Affolder {\it et al.}  [CDF Collaboration],
Phys.\ Rev.\ Lett.\  {\bf 85}, 2886 (2000);

D.~Acosta {\it et al.}  [CDF Collaboration],
Phys.\ Rev.\ Lett.\  {\bf 88}, 161802 (2002).





\bibitem{VHBT}
M.~Vanttinen, P.~Hoyer, S.~J.~Brodsky and W.~K.~Tang,
Phys.\ Rev.\ D {\bf 51}, 3332 (1995).


\bibitem{lemo}
Y.~Lemoigne et al. (WAII Collaboration), Phys. Lett. D {\bf 113B},
509 (1982);

L.~Antoniazzi et al. (E705 Collaboration), Phys. Rev.Lett {\bf
70}, 383 (1993).


\bibitem{BI}
B.~L.~Ioffe,
Phys.\ Rev.\ Lett.\  {\bf 39}, 1589 (1977); Phys.\ Atom.\ Nucl.\
{\bf 57}, 1705 (1994).




\bibitem{data}


J.~Badier {\it et al.}  [NA3 Collaboration],
Z.\ Phys.\ C {\bf 20}, 101 (1983).


C.~Biino {\it et al.},
Phys.\ Rev.\ Lett.\  {\bf 58}, 2523 (1987).

\bibitem{BLKh} B.L.Ioffe, D.E.Kharzeev, hep-ph/0306062,

\bibitem{BLDE} B.L.Ioffe, D.E.Kharzeev, hep-ph/0306176.





\end{thebibliography}
\end{document}